\documentclass[aps,pre,showpacs,twocolumn,groupedaddress,superscriptaddress]{revtex4}
\usepackage{graphicx}%
\usepackage{amsmath}
\usepackage{dcolumn}
\usepackage{epsfig}
\usepackage{graphics}
\usepackage{bm}

\begin{document}

\title{Mass synchronization: Occurrence and its control with possible applications to brain dynamics}

\author{V.~K.~Chandrasekar}%
 \email{chandru25nld@gmail.com}
\affiliation{Centre for Nonlinear Dynamics, School of Physics,
Bharathidasan University, Tiruchirappalli - 620 024, Tamilnadu, India}

\author{Jane H.~Sheeba}%
 \email{jane.sheeba@gmail.com}
\affiliation{Centre for Nonlinear Dynamics, School of Physics,
Bharathidasan University, Tiruchirappalli - 620 024, Tamilnadu, India}

\author{M.~Lakshmanan}%
 \email{lakshman@cnld.bdu.ac.in}
\affiliation{Centre for Nonlinear Dynamics, School of Physics,
Bharathidasan University, Tiruchirappalli - 620 024, Tamilnadu, India}

\date{\today}

\begin{abstract}
Occurrence of strong or mass synchronization of a large number of neuronal populations in the brain characterizes its pathological states. In order to establish an understanding of the mechanism underlying such pathological synchronization we present a model of coupled populations of phase oscillators representing the interacting neuronal populations. Through numerical analysis, we discuss the occurrence of mass synchronization in the model, where a source population which gets strongly synchronized drives the target populations onto mass synchronization. We hypothesize and identify a possible cause for the occurrence of such a synchronization, which is so far unknown: Pathological synchronization is caused not just because of the increase in the strength of coupling between the populations but also because of the strength of the strong synchronization of the drive population. We propose a demand-controlled method to control this pathological synchronization by providing a delayed feedback where the strength and frequency of the synchronization determines the strength and the time delay of the feedback. We provide an analytical explanation for the occurrence of pathological synchronization and its control in the thermodynamic limit.
\end{abstract}



\maketitle

\textbf{Strong synchronization of several groups of neuronal oscillators occurs during pathological states like Parkinson's tremor, epileptic seizures, etc. This synchronization first occurs locally in a particular region of the brain due to synchronization of a group of neurons. This group then acts as a pacemaker triggering synchronization to various other neuronal populations resulting in a mass synchronization. Deep brain stimulation is a standard therapy for patients with this disorder. However since the actual cause and mechanism of the occurrence of mass synchronization in the brain is not yet fully known, there is always a need to understand the same in order to be able to control it effectively. In this article we attempt to gain this understanding by studying a system of coupled populations of phase oscillators representing interacting neuronal populations in the brain. We find the occurrence of mass synchronization in such a system where the occurrence of synchronization in the source population triggers synchronization in the target populations also. We hypothesize a possible cause for the occurrence of mass synchronization in the brain, which is so far not known. We propose a demand controlled method to control this mass synchronization by providing a delayed feedback. We provide numerical evidence and analytical explanation for the occurrence and control of neuronal mass synchronization. }

\section{Introduction}

Synchronization is an ubiquitous phenomenon and a topic of active research in recent years~\cite{Winfree:67,Kuramoto:84,Pikovsky:01}. Given that it is not always a desirable phenomenon, synchrony control mechanism are always required. For example, in the brain, when synchronization supports cognition via temporal coding of information \cite{Singer:99,Jane:08b}, it is desirable while at the same time it is undesirable when synchronization of a mass of neuronal oscillators occurs at a
particular frequency band resulting in pathologies like trauma, Parkinson's tremor and so on. Other examples include lasers and Josephson junction arrays \cite{Trees:05}, emission of microwave frequencies by coupled spin torque nano oscillators \cite{Mohanty:05} where synchronization is desirable, while in the cases of epileptic seizures \cite{Timmermann:03}, Parkinson's tremor \cite{Percha:05}, event related desynchronization \cite{Pfurtschelle:99,Jane:09}, or pedestrians on the Millennium Bridge \cite{Strogatz:01}, it is undesirable.

Pathological strong (or mass) synchronization occurs when several groups of neurons enter into synchronization and the strength of the synchronization is large because of the large number of neuronal networks that are engaged in synchrony. In particular, tremors in the case of Parkinson's disease (PD), essential tremor and several other neurological diseases like epileptic seizures are caused by mass synchronization of oscillating neuronal networks. Synchronization in such cases may also be called abnormal synchronization, since under normal conditions mass synchronization does not occur. Clusters of neurons fire in a synchronized manner with a frequency similar to that of the tremor and act like a pacemaker driving the pre-motor and motor cortical areas of the brain causing mass synchronization \cite{Alberts:69,Volkmann:04}. In fact it has been shown that the local field potentials in the thalamic and the sub-thalamic nucleus (STN) drive the tremor in the limb \cite{Levy:00,Smirnov:08,Reck:09}. In the case of healthy subjects these neurons in the STN oscillate in a desynchronized manner \cite{Nini:95}. Deep brain stimulation is a standard therapy for such patients with such tremors where a train of strong high frequency electric pulse is administered to the areas like STN where there is mass synchronization. In spite of the great amount of interest and research focussed on pathological tremors, the actual mechanism that cause these tremors are still unknown \cite{Mormann:00,Hammond:07,Andrzejak:09,Timmermann:09,Kane:09}. Further, controlling of pathological tremors is also a problem which has recently gained much attention \cite{Rosenblum:04,Popovych:05,Cilia:09,Ray:09}.

Given that the interaction mechanism and the topological structure of the neuronal networks are highly complex and are not completely known, and that there is a lack of technology to measure synchronization in every region of neuronal network, it is highly complicated to gain a complete understanding of the underlying mechanism in these pathological states. In fact, the actual mechanism for synchronous oscillations in Parkinson's disease remains unknown. Nevertheless, one can always represent the system of interacting neurons as a mathematical model and simulate the challenging case of a pathological synchronization to understand the underlying dynamical aspects. With this motive, in this paper, we hypothesize the possible cause for the occurrence of mass synchronization in PD tremors and also provide a method to control it.

The plan of the paper is as follows: In the following Sec. \ref{model} we introduce a model of coupled populations of phase oscillators which represents interacting populations of neurons. We present our numerical finding of the occurrence of mass synchronization where the source population drives the target in Sec. \ref{mass}. In Sec. \ref{control} we propose a method to control the occurrence of pathological synchronization by providing a delayed feedback. We analytically explain the occurrence of mass synchronization and the method to control it in Sec. \ref{anal}. Finally in Sec. \ref{conc} we present our conclusions.

\section{The model}
\label{model}
 We consider a model of interacting populations of neurons which is represented by populations of coupled phase oscillators. One of the populations is considered to be a source which acts as a pacemaker and possibly drives the other populations (named as targets) onto mass synchronization. The model equations read as
\begin{eqnarray}
\label{cho01}
\dot{\theta_i}^{(\eta)}= \omega_i^{(\eta)} - \sum_{\eta'=1}^{N'}\frac{\sigma_{\eta\eta'}}{N_{\eta'}}\sum_{j=1}^{N_{\eta'}=N}
\sin(\theta_i^{(\eta)}-\theta_j^{(\eta')}+\alpha_{\eta\eta'}),\nonumber\\
\qquad\qquad\qquad\qquad\qquad\qquad\qquad i=1,2, \ldots, N_{\eta'}=N,
\end{eqnarray}
where $\eta=1,2,\ldots N'$, $N'$ is the number of populations and $N_{\eta'}$ is the number of neurons in population $\eta'$ which is assumed to be the same in all the populations, that is equal to $N$. $\sigma_{\eta\eta'}$ is the strength of the coupling between the neurons in $\eta'$ and those in $\eta$. Note that $\eta=\eta' (1,2,\ldots N')$ corresponds to interconnection among the neurons of the same population. Here $\omega_k^{(\eta)}$ is the natural frequency of the $k$th neuron in the population $\eta$  and $0 \leq |\alpha_{\eta\eta'}| <\pi/2$ is the phase lag. In reality every oscillator has different $\alpha$, but we have taken them to be the same for all of them, for simplicity.

\begin{figure}
\begin{center}
\includegraphics[width=7.50cm]{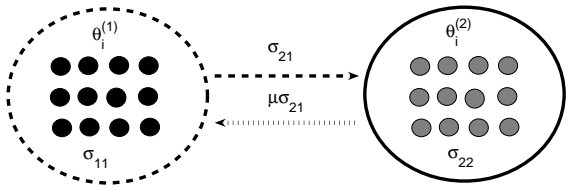}
\caption{Schematic representation of system (\ref{cho01}) for $N'=2$. $\theta_i^{(1)}$ is the source population and $\theta_i^{(2)}$ is the target. The coupling strengths within the populations are quantified by the parameters $\sigma_{11}$ and $\sigma_{22}$. The coupling strengths from the source to the target and the target to the source are quantified by the parameters $\sigma_{12}=\mu\sigma_{21}$ and $\sigma_{21}$, respectively. }
\label{new_cfi_schm1}
\end{center}
\end{figure}

For a better understanding, a schematic diagram of the model is given in Fig. \ref{new_cfi_schm1} for $N'=2$, where the source population is represented by a dotted circle ($\theta_i^{(1)}$) and the target is represented by a solid circle ($\theta_i^{(2)}$). In normal states it is the low frequency signal that drives the high frequency signal \cite{Musizza:07,Marshall:06}, but in the case of pathological synchronization, high frequency signal from the source drives the target populations that are oscillating at (relatively) low frequencies on to synchronization. Therefore we have chosen $\sigma_{12}=\mu\sigma_{21}$, where $0\leq\mu<1$, since
the coupling from the source to the target ($\sigma_{21}$) is stronger
than the coupling from the target to the source ($\sigma_{12}$) in pathological states. Oscillators inside the source and the target populations are coupled among themselves with strengths $\sigma_{11}$ and $\sigma_{22}$, respectively. In order to quantify synchronization within a population, we can define the order parameter as
\begin{eqnarray}
\label{cho01a}
z_{\eta}=r_{\eta}e^{i\psi^{(\eta)}}= \frac{1}{N}\sum_{j=1}^Ne^{i\theta_j^{(\eta)}}.
\end{eqnarray}
When $r_{\eta}=1$, there is complete synchronization within the $\eta$ population since in this state the phases of all the oscillators in that population are the same. When $r_{\eta}$ takes values in between $0$ and $1$ there is partial  desynchronization and when $r_{\eta}=0$ there is complete desynchronization in the $\eta$ population. The time average of $r_{\eta}$ is given by
\begin{eqnarray}
\label{cho01b}
R_{\eta}=<r_{\eta}>=\frac{1}{T}\int_{0}^{T}r_{\eta}dt,
\end{eqnarray}
which is used to characterize the occurrence of strong synchronization in a population. Numerically, strong synchronization is characterized by $R_\eta>0.8$ and desynchronization is characterized by $R_\eta\leq0.3$ for $T=10^5$ time units. When strong synchronization occurs in all the populations due to the strong coupling strength in the source it is called mass synchronization.
From equations (\ref{cho01}) and (\ref{cho01a}) the model equations can be rewritten as
\begin{eqnarray}
\label{cho02}
\dot{\theta_i}^{(\eta)}= \omega_i^{(\eta)} - \sum_{\eta'=1}^{N'}\sigma_{\eta\eta'}r_{\eta'}
\sin(\theta_i^{(\eta)}-\psi^{(\eta')}+\alpha_{\eta\eta'}).
\end{eqnarray}
In order to find out the cause for the occurrence of mass synchronization and also to find ways to control it, we numerically simulate system (\ref{cho02}) using Runge-Kutta fourth order routine with a time step of $0.01$. We have set $N=1000$ and have assumed a Lorentzian distribution for the oscillator frequencies given by
\begin{eqnarray}
\label{cho10}
g(\omega^{(\eta)})&=&\frac{\gamma_{\eta}}{\pi}\bigg[(\omega^{(\eta)}-\omega_{\eta})^2+\gamma_{\eta}\bigg]^{-1},
\end{eqnarray}
where $\gamma$ is the half width at half maximum and $\omega_{\eta}$ is the central frequency. We have discarded transients of the order of $10^5$ in our simulations. Simulations lasted for $2\times10^5$ steps and the results are shown for a small window of time (which we label to start with 0 in the figures for convenience) towards the end of the total simulation time. The initial phases are considered to be randomly distributed between 0 and 2$\pi$. In the following sections we will discuss how the occurrence of synchronization in the source population causes mass synchronization in the target populations and will also present a method to control this mass synchronization.

The model we consider here is of mass model type and focusses on the macroscopic properties such as the synchronization of large population of neurons. The model therefore does not include a detailed microscopic description to incorporate the effects of hierarchal connections, spiking activities of individual neurons and populations, the role of neurotransmitters and so on \cite{Diesmann:99,van Rossum:02,David:05,Harrison:05,Kumar:10}. We make use of the macroscopic modeling approach to strike a balance between abstract and detailed modeling while keeping the number of variables and parameters to a reasonable minimum. Thus we attempt to make a qualitative concurrence between the model results and the occurrence of pathological mass synchronization.

\section{Occurrence of mass synchronization}
\label{mass}
\begin{figure}
\begin{center}
\includegraphics[width=7.50cm]{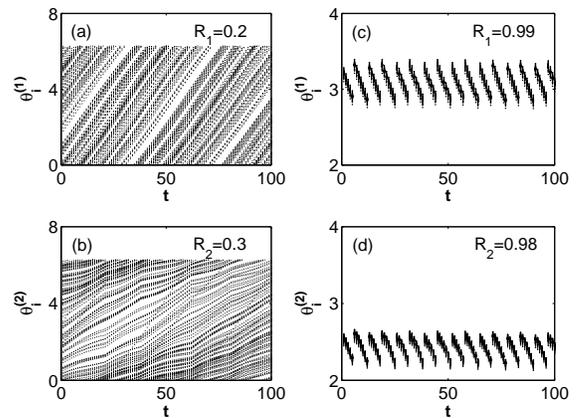}
\caption{The time evolution of the phases in the source (panels (a) and (c)) and the target (panels (b) and (d)) populations are plotted for $N=1000$. Panels (a) and (b) represent the initial desynchronized state characterized by $R_1=0.2$ and $R_2=0.3$ and panels (c) and (d) represent the occurrence of strong synchronization in the target $\theta_i^{(2)}$ ($R_2=0.98$) when the source population $\theta_i^{(1)}$ gets synchronized ($R_1=0.99$). Here $\sigma_{22}=0.1$, $\sigma_{21}=1.5$, $\mu=0.3$, $\gamma_{1,2}=0.05$, $\omega_{1}=1.5$, $\omega_2=0.5$, $\alpha_{ij}=0$, $i,j=1,2$, and $\sigma_{11}=0.1$ in (a) and (b) and $\sigma_{11}=2.5$ in (c) and (d).}
\label{new_cfi_fig1}
\end{center}
\end{figure}

\begin{figure}
\begin{center}
\includegraphics[width=8.50cm]{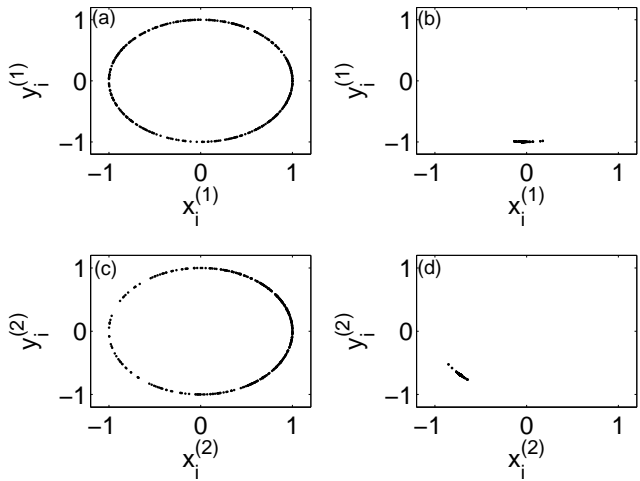}
\caption{(a)-(d) are the snapshots of the phase portraits on the ($x_i$,$y_i$) plane corresponding to Fig. \ref{new_cfi_fig1} (a)-(d), where $x_i=\cos\theta_i$ and $y_i=\sin\theta_i$.}
\label{new_kur_xy1}
\end{center}
\end{figure}

In order to illustrate the occurrence of mass synchronization, we have taken $N'=2$, so that we have a system of one target and one source populations, each having 1000 oscillators. We have verified our results for the $N'=3$ case also (see Fig. \ref{cfi_Rplots1} below). To start with, we consider a state when both the source and the target populations are desynchronized ($\sigma_{11}=\sigma_{22}=0.1$) with a very low value of the average order parameters $R_1=0.2$ and $R_2=0.3$, even with a strong coupling between the source and the target ($\sigma_{21}=1.5$ and $\mu=0.3$). Upon increasing the coupling strength within the source population, that is $\sigma_{11}=2.5$, we find that strong synchronization occurs in the source (characterized by $R_1=0.99$) which in turn induces strong synchronization in the target (characterized by $R_2=0.98$). Note that this strong synchronization occurs in the target due to its coupling with the source, that is $\sigma_{21}$, and not because of the coupling within itself, since $\sigma_{22}=0.1$ only. This can be seen from Fig. \ref{new_cfi_fig1} where we have plotted the time evolution of the phases in the source (panels (a) and (c)) and the target (panels (b) and (d)) populations. Panels (a) and (b) show the initial desynchronization state and (c) and (d) depict the occurrence of mass synchronization due to the increase in the coupling $(\sigma_{11})$ within the source population. Fig. \ref{new_kur_xy1} shows the  snapshots of the distribution of the phases in the ($x_i,y_i$)=($\cos\theta_i, \sin\theta_i$) plane where the panels (a)-(d) correspond to those in Fig. \ref{new_cfi_fig1}. In panels (a) and (b) the oscillators in the source and the target populations are desynchronized initially and hence the phases are distributed throughout the unit circle. Panels (c) and (d) depict the distribution of phases in the source and the target populations, respectively, in the mass synchronization state; the phases occupy a rather narrow region on the unit circle.

Thus it becomes evident that for the onset of pathological synchronization the neurons in the source population are strongly coupled among themselves. Due to this, the occurrence of a strong synchronization in the source population is propagated onto the target populations because of an increase in the order parameter $r_1$ of the source, see equation (\ref{cho02}). As a consequence $\sigma_{21}r_1$ increases faster than $\mu\sigma_{21}r_2$ and hence the source drives the target and makes it to get synchronized with itself. Under normal conditions, however $\sigma_{11}$ does not attain higher values to cause strong synchronization in the source. Thus for the occurrence of mass synchronization not only the coupling strength between the populations is crucial but also the strength of the synchronization of the source (characterized by $r_1$ or $R_1$) plays a vital role. This is evident from Fig. \ref{cfi_Rplots1} where we have plotted the time average of $r_\eta$ $(R_\eta)$ for varying coupling strength $\sigma_{11}$ of the source for the two cases, $N'=2$ and $N'=3$.

For the case $N'=3$ we have considered three populations of coupled phase oscillators each having 1000 oscillators with $\sigma_{\eta\eta}=0.05$, $\sigma_{\eta1}=1.5$, $\sigma_{1\eta}=\mu\sigma_{\eta1}$, for $\eta=2,3$, $\mu=0.2$, $\gamma_{1,2,3}=0.05$, $\omega_{1}=1.5$, $\omega_{2,3}=0.5$, and $\sigma_{23}=\sigma_{32}=0.01$. One of the populations is the source (with average order parameter $R_1$) and there are two target populations (with average order parameters $R_2$ and $R_3$) in the system. For both the cases $N'=2$ and $N'=3$, as $\sigma_{11}$ increases the strength of the synchronization of the source increases and as a consequence synchronization in the target populations also increases. Interestingly, as one can see from the figure, for a given $\sigma_{11}$, the target population achieves strong synchronization with the source even before the source population achieves complete synchronization. Our model results prove to be valid for a system of large number of coupled populations also and we have presented here the case of three populations for illustration.
\begin{figure}
\begin{center}
\includegraphics[width=7.50cm]{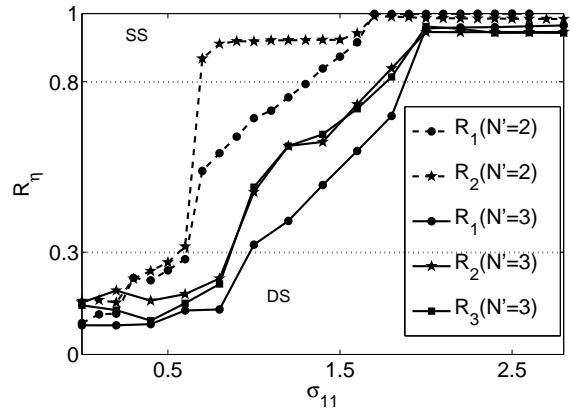}
\caption{The time average of $r_\eta$ for increasing $\sigma_{11}$ is plotted for two cases, $N'=2$ (dotted lines) and $N'=3$ (solid lines). Here $\sigma_{\eta\eta}=0.05$, $\sigma_{\eta1}=1.5$, $\sigma_{1\eta}=\mu\sigma_{\eta1}$, for $\eta=2,3$, $\mu=0.2$, $\alpha_{ij}=\pi/2-0.1$, $i,j=1,2,3$, $\gamma_{1,2,3}=0.05$, $\omega_{1}=1.5$, $\omega_{2,3}=0.5$, and $\sigma_{23}=\sigma_{32}=0.01$. The regions DS and SS denote the desynchronization and strong synchronization states, respectively. }
\label{cfi_Rplots1}
\end{center}
\end{figure}
\section{Method to control mass synchronization}
\label{control}
In order to control the occurrence of mass neuronal synchronization we apply a delayed feedback force to the populations in which case the model equation (\ref{cho01}) becomes
\begin{eqnarray}
\label{cho02a}
\dot{\theta_i}^{(\eta)}&=& \omega_i^{(\eta)} - \sum_{\eta'=1}^{N'}\sigma_{\eta\eta'}r_{\eta'}
\sin(\theta_i^{(\eta)}-\psi^{(\eta')}+\alpha_{\eta\eta'})\nonumber\\
&&\pm FR_{\tau}\sin(\theta_i^{(\eta)}-\phi_{\tau}),\;\;i=1,2, \ldots, N,
\end{eqnarray}
where $R_\tau=R(t-\tau)$ and $\phi_{\tau}=\phi(t-\tau)$ and $Z(t)=\sum_{\eta'=1}^{N'}z_{\eta'}=Re^{i\phi}=\frac{1}{N'}\sum_{\eta'=1}^{N'}r_{\eta'}e^{i\psi_{\eta'}}$ is the global order parameter. The feedback can be either positive or negative as denoted by the $\pm$ sign in the above equation since the dynamics of the system does not change qualitatively upon using either $+$ or $-$ signs \cite{Sheeba:09a}. Also, one can note that the $\pm$ sign can be interchanged upon replacing $\phi$ with $\phi-\pi$.

\begin{figure}
\begin{center}
\includegraphics[width=7.50cm]{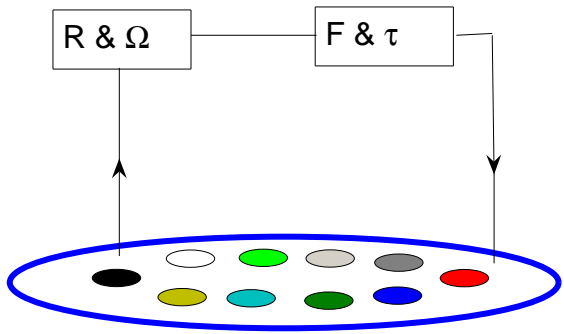}
\caption{(Color online) Schematic representation of the control set up. Filled circles represent the neuronal populations. The strength of the global synchronization $(R)$ and the corresponding frequency $(\Omega)$ are measured. A delayed feedback of strength $F$ and a time delay of $\tau$ is fed back to the neuronal populations in order to cause desynchronization. }
\label{new_cfi_schm2}
\end{center}
\end{figure}

The global order parameter $R(t)$ measures the strength of synchronization of all the populations (combined) in the system while the order parameter $r_\eta$ measures the strength of synchronization of the $\eta$th population.  The schematic representation of this mass synchronization control set up is shown in Fig. \ref{new_cfi_schm2}. The strength of the global synchronization $R(t)$ and the corresponding frequency $\Omega \;(=d\phi/dt)$ are measured from the populations of neurons denoted by filled circles. A delayed feedback of strength $F$ and a time delay of $\tau$ is fed back to the neuronal populations in order to control mass synchronization. This method is effective since one can have control over certain parameters like time delay and the strength of the feedback. There is also no need to know about the interactions among the neurons or the characteristics of individual neuronal oscillators, but all we need is the macroscopic information. Further, the feedback is demand controlled because $F$ and $\tau$ are fixed depending upon the values of $R$ and $\Omega$. In addition, there is no need to seek for a new controlling signal that stimulates the neuronal populations, but the measured pathological signal is used to control the same. We choose to provide delayed feedback because it can affect the activity of the individual neurons by shifting their phases.

\begin{figure}
\begin{center}
\includegraphics[width=7.50cm]{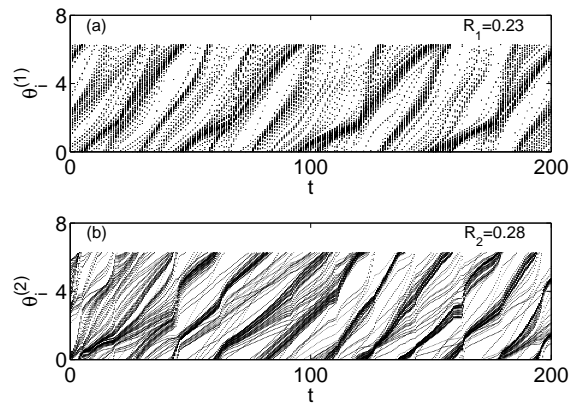}
\caption{(a) and (b) show the occurrence of desynchronization for $\tau=1.0$ and $F=5.5$ in the source ($\theta_i^{(1)}$) and the target ($\theta_i^{(2)}$) populations, respectively due to delayed feedback, following panels (c) and (d) of Fig. \ref{new_cfi_fig1}. Here N=256.}
\label{new_cfi_fig2}
\end{center}
\end{figure}

\begin{figure}
\begin{center}
\includegraphics[width=7.50cm]{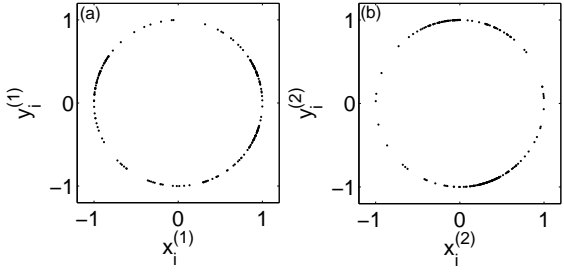}
\caption{(a) and (b) are the snapshots of the phase portraits on the ($x$,$y$) plane corresponding to Fig. \ref{new_cfi_fig2} (a) and (b) for N=256.}
\label{delay_kur_xy1}
\end{center}
\end{figure}

The numerical simulation of equation (\ref{cho02a}) showing the control of mass synchronization is given in Fig. \ref{new_cfi_fig2} which follows the state of mass synchronization shown in Fig. \ref{new_cfi_fig1} (panels (c) and (d)) after applying the delayed feedback ($\tau=1.0$ and $F=5.5$). Due to the delayed feedback both the source (panel (a)) and the target (panel (b)) populations are desynchronized. The corresponding distribution of phases on the ($x_i,y_i$) plane is shown in Fig. \ref{delay_kur_xy1} where the phases are distributed throughout the unit circle denoting the desynchronization states of the source (panel (a)) and the target (panel (b)) populations. For numerical simulations including delayed feedback we have set $N=256$. We have also checked that the results are size independent by calculating them for $N=$128 and 512.

In the following section we analytically show the occurrence of mass synchronization and its control with delayed feedback in support of the above numerical findings as follows: (i) We will show that when the source is desynchronized the target population will also be desynchronized. (ii) Then when strong synchronization occurs in the source the same is induced in the target also. (iii) In the strong synchronization state, when delayed feed back is introduced the synchronization is broken for suitable values of the strength of the feedback and the value of the time delay.
\section{Analytical explanation}
\label{anal}
In order to analytically explain the occurrence of mass synchronization and its control, we analyze system (\ref{cho01}) in the continuum limit $N\rightarrow \infty$.  In this limit, a probability density for the oscillator phases can be defined as $\rho^{(\eta)}(\omega,\theta,t)$, which describes the number of oscillators with phases within $[\theta,\theta+d\theta]$ and natural frequencies between $[\omega$,$\omega+d\omega$] at time $t$. This distribution $\rho^{(\eta)}(\omega,\theta,t)$ obeys the evolution equation
\begin{eqnarray}
\label{cho03}
\frac{\partial\rho^{(\eta)}}{\partial
t}=-\frac{\partial}{\partial\theta}(\rho^{(\eta)}v^{(\eta)}), \eta=1,2, \ldots, N'
\end{eqnarray}
where from equation (\ref{cho02}) (in the continuum limit) the velocity $v^{(\eta)}$ is given by
\begin{eqnarray}
\label{cho04}
\small
v^{(\eta)}&=&\omega^{(\eta)} - \sum_{\eta'=1}^{N'}\sigma_{\eta\eta'}\int_{-\infty}^{\infty}\int_0^{2\pi}
\sin(\theta^{(\eta)}-\theta^{'(\eta')}+\alpha_{\eta\eta'})\nonumber \\&&\times \rho^{(\eta')}(\omega^{(\eta')},\theta^{'(\eta')},t)d\theta^{'(\eta')}d\omega^{(\eta')},
\nonumber \\&&
\qquad\qquad\qquad\qquad\qquad \eta=1,2, \ldots, N'.
\end{eqnarray}

Also, in the continuum limit the order parameter (\ref{cho01a}) for the population $\eta$ becomes
\begin{eqnarray}
\label{cho05}
z_{\eta}&=&\int_{-\infty}^{\infty}\int_0^{2\pi}e^{i\theta^{(\eta)}}
\rho^{(\eta)}(\omega^{(\eta)},\theta^{(\eta)},t)d\theta^{(\eta)}d\omega^{(\eta)}.
\end{eqnarray}
Now we rewrite the velocity in terms of the order parameters to get
\begin{eqnarray}
\label{cho06}
\small
v^{(\eta)}&=&\omega^{(\eta)} - \sum_{\eta'=1}^{N'}\frac{\sigma_{\eta\eta'}}{2i}\bigg(e^{i(\theta^{(\eta)}+\alpha_{\eta\eta'})}z_{\eta'}^{\ast}
\nonumber\\ &&-e^{-i(\theta^{(\eta)}+\alpha_{\eta\eta'})}z_{\eta'}\bigg).
\end{eqnarray}
Upon introducing the ansatz of Ott and Antonsen \cite{Ott:08} we consider a special class of the density functions in the form of Poisson kernel as
\begin{eqnarray}
\label{cho07}
\rho^{(\eta)}(\omega^{(\eta)},\theta^{(\eta)},t)&=&\frac{g(\omega^{(\eta)})}{2\pi} \bigg[1+\bigg( \sum_{k=1}^{\infty}(a_{\eta}(\omega^{(\eta)},t)e^{i\theta^{(\eta)}})^k\nonumber\\
&&+c.c\bigg)\bigg], \eta=1,2, \ldots, N',
\end{eqnarray}
where $g(\omega^{(\eta)}$ is the distribution of the oscillator frequencies.
Here c.c denotes the complex conjugate of the preceding term. The main advantage of using this ansatz is that it has the same function $a_{\eta}$ in all the Fourier harmonics, except that $a_{\eta}$ is raised to the $n$th power in the $n$th harmonic. Thus one can reduce the infinite set of dynamical equations exactly into a set of finite number of equations. Using the above expression (\ref{cho07}) for $\rho^{(\eta)}$ into the evolution equation (\ref{cho03}), we get the equation for $a_{\eta}$ as
\begin{eqnarray}
\label{cho08}
\dot{a}_{\eta}&+&i\omega^{(\eta)}a_{\eta}
\nonumber\\&=&\sum_{\eta'=1}^{N'}\frac{\sigma_{\eta\eta'}}{2}
\bigg(e^{i\alpha_{\eta\eta'}}z_{\eta'}^{\ast}-e^{-i\alpha_{\eta\eta'}}z_{\eta'}a_{\eta}^2\bigg),\nonumber\\&&\qquad\qquad\qquad\qquad\qquad
\eta=1,2, \ldots, N'.
\end{eqnarray}
Now, substituting the Poisson kernel (\ref{cho07}) into the order parameter equation (\ref{cho05}) we get
\begin{eqnarray}
\label{cho09}
z_{\eta}&=&\int_{-\infty}^{\infty}a_{\eta}(\omega^{(\eta)},t)g(\omega^{(\eta)})d\omega^{(\eta)}.
\end{eqnarray}
Assuming a Lorentzian distribution (\ref{cho10}) for $g(\omega^{(\eta)})$ and evaluating the integral in equation (\ref{cho09}) by contour integration we get $z_{\eta}^{\ast}= a_{\eta}(\omega_{\eta}-i\gamma_{\eta},t)$. Thus equation (\ref{cho08}) becomes
\begin{eqnarray}
\label{cho11}
\dot{z}_{\eta}&+&(\gamma_{\eta}-i\omega_{\eta})z_{\eta}
\nonumber\\&=&\sum_{\eta'=1}^{N'}\frac{\sigma_{\eta\eta'}}{2}
\bigg(e^{-i\alpha_{\eta\eta'}}z_{\eta'}-e^{i\alpha_{\eta\eta'}}z_{\eta'}^{\ast}z_{\eta}^2\bigg),\nonumber\\&&\qquad\qquad\qquad\qquad\qquad
\eta=1,2, \ldots, N'.
\end{eqnarray}
Now we express the amplitude equation (\ref{cho11}) in polar coordinates $z_{\eta}= r_{\eta}e^{\psi_{\eta}}$ as (see equation (\ref{cho01a}))
\begin{eqnarray}
\label{cho12}
\dot{r}_{\eta}&=&-\gamma_{\eta}r_{\eta}+(\frac{1-r_{\eta}^2}{2})\sum_{\eta'=1}^{N'}\sigma_{\eta\eta'}r_{\eta'}
\cos(\psi_{\eta}-\psi_{\eta'}+\alpha_{\eta\eta'}),\nonumber\\
\dot{\psi}_{\eta}&=&\omega_{\eta}-(\frac{1+r_{\eta}^2}{2r_{\eta}})\sum_{\eta'=1}^{N'}\sigma_{\eta\eta'}r_{\eta'}
\sin(\psi_{\eta}-\psi_{\eta'}+\alpha_{\eta\eta'}),
\nonumber\\&&\qquad\qquad\qquad\qquad\qquad
\eta=1,2, \ldots, N'.
\end{eqnarray}
Using these equations we will explain the occurrence of pathological synchronization when the source is synchronized, and its control by applying delayed feedback, in the following subsections.

\subsection{Occurrence of mass synchronization}
As noted earlier in Sec. III, we have numerically found the occurrence of mass synchronization for the case $N'=2$ (as well as $N'=3$) in Eq. (\ref{cho01}). In order to explain this analytically we rewrite the amplitude equation (\ref{cho12}) for the case $N'=2$ as
\begin{eqnarray}
\label{cho13}
\dot{r}_{1}&=&-\gamma_{1}r_{1}+(\frac{1-r_{1}^2}{2})(\sigma_{11}r_{1}\cos(\alpha_{11})
\nonumber\\&&+\mu\sigma_{21}r_2
\cos(\psi_{1}-\psi_{2}+\alpha_{12})),\nonumber\\
\dot{\psi}_{1}&=&\omega_{1}-(\frac{1+r_{1}^2}{2r_{1}})(\sigma_{11}r_{1}\sin(\alpha_{11})
\nonumber\\&&+\mu\sigma_{21}r_2
\sin(\psi_{1}-\psi_{2}+\alpha_{12})),\\
\dot{r}_{2}&=&-\gamma_{2}r_{2}+(\frac{1-r_{2}^2}{2})(\sigma_{22}r_{2}\cos(\alpha_{22})
\nonumber\\&&+\sigma_{21}r_1 \cos(\psi_{2}-\psi_{1}+\alpha_{21})),
\nonumber\\
\dot{\psi}_{2}&=&\omega_{2}-(\frac{1+r_{2}^2}{2r_{2}})(\sigma_{22}r_{2}\sin(\alpha_{22})
\nonumber\\&&+\sigma_{21}r_1
\sin(\psi_{2}-\psi_{1}+\alpha_{21})).\label{cho13a}
\end{eqnarray}
To start with we will consider the case $\mu=0$ in equations (\ref{cho13}) and (\ref{cho13a}) for analytical simplicity to identify pathological synchronization in neuronal populations so that the source strongly (completely) drives the target populations. However, in general one can observe the occurrence of pathological synchronization for $\mu<1$ also as in the case of numerical simulations where $\mu=$0.2 and 0.3. For this purpose, we will also consider below the case $\mu\neq 0$ in equations (\ref{cho13}) and (\ref{cho13a}) so as to provide a theoretical basis for the numerical findings. Consequently, we can conclude that when the strength of the synchronization increases in the source it induces synchronization in the target.

From equation (\ref{cho13}), for $\mu=0$, the synchronization of the source is characterized by the stability of the fixed point $r_1^s=\sqrt{1-2\gamma_1/\bar{\sigma}_{11}}$, where $\bar{\sigma}=\sigma \cos(\alpha)$. On the other hand, the desynchronization state is characterized by the stability of the fixed point $r_1^d=0$. When $\bar{\sigma}_{11}<2\gamma_1$, the fixed point $r_1^d$ becomes stable and there is no synchronization in the source. In this state the equation for the target population is given as
\begin{eqnarray}
\dot{r}_{2}&=&-\gamma_{2}r_{2}+(\frac{1-r_{2}^2}{2})(\sigma_{22}r_{2}\cos(\alpha_{22})
\nonumber\\
\dot{\psi}_{2}&=&\omega_{2}-(\frac{1+r_{2}^2}{2r_{2}})\sigma_{22}r_{2}\sin(\alpha_{22}).\label{cho13aa}
\end{eqnarray}
Again one can check that for $\bar{\sigma}_{22}<2\gamma_2$ the fixed point $r_2=0$ is stable and the target population is desynchronized. Thus when the source is desynchronized, the target population is also desynchronized. This result holds good for $\mu<1$ also as is evident from our numerical findings as depicted in panels (a) and (b) of Figs. \ref{new_cfi_fig1} and \ref{new_kur_xy1} for $\mu=0.3$.

On the other hand, when the coupling strength in the source $\bar{\sigma}_{11}$ increases so that $\bar{\sigma}_{11}>2\gamma_1$ the fixed point $r_1^d$ becomes unstable and $r_1^s$ becomes stable thus establishing synchronization in the source. The synchronization strength of the source increases as $\sqrt{1-2\gamma_1/\bar{\sigma}_{11}}$. After synchronization in the source is established, equation (\ref{cho13a}) reduces to
\begin{eqnarray}
\label{cho14}
\dot{r}_2&=&-\gamma_{2}r_2+(\frac{1-r_2^2}{2})(\sigma_{22}r_2\cos(\alpha_{22})
\nonumber\\&&+\sigma_{21}\sqrt{1-2\gamma_1/\bar{\sigma}_{11}}
\cos(\psi+\alpha_{21})),\nonumber\\
\dot{\psi}&=&\bar{\omega}-(\frac{1+r_2^2}{2r_2})(\sigma_{22}r_2\sin(\alpha_{22})
\nonumber\\&&+\sigma_{21}\sqrt{1-2\gamma_1/\bar{\sigma}_{11}}
\sin(\psi+\alpha_{21})),
\end{eqnarray}
where $\psi=\psi_2-\psi_1$ and $\bar{\omega}=\omega_{2}-\omega_{1}+(\sigma_{11}-\gamma_1)\tan(\alpha_{11})$. This equation does not admit the fixed point $r_2=0$. This means that when synchronization emerges in the source, the target population also begins to get synchronized. The strength of the synchronization in the target increases according to $\sigma_{21}\sqrt{1-2\gamma_1/\bar{\sigma}_{11}}$, eventually leading to synchronization with the source (we note here that in \cite{Ott:08,Childs:08} an
equation similar to (\ref{cho14})  has been analyzed in connection with the stability of the synchronized and desynchronized states). This results in the pathological strong synchronization which corresponds to our numerical findings as depicted in panels (c) and (d) of Figs. \ref{new_cfi_fig1} and \ref{new_kur_xy1}.

Even for $\mu\neq 0$, from equations (\ref{cho13}) and (\ref{cho13a}), we can establish that when the source is completely desynchronized the target is also completely desynchronized. That is when the fixed point $r_1=0$ exists, $r_2=0$ also exists. When $\sigma_{11}$ increases, synchronization is induced in the source. Hence, the fixed point $r_1=0$ does not exist as a result of which $r_2=0$ also becomes nonexistent. This means that when synchronization arises in the source, the target also spontaneously becomes synchronized. This can be described as follows: The eigenvalue of the fixed point $(r_1,r_2)=(0,0)$ is
\begin{eqnarray}
\lambda_{\pm} =
-\gamma+\frac{\kappa}{4}\pm\frac{1}{2}(\xi
-i\hat{A}\Delta\omega-\Delta\omega^2)^{\frac{1}{2}}
-i\bar{\omega},
 \label{tany06}
\end{eqnarray}
where $\gamma_{1,2}=\gamma$, $\alpha_{i,j}=0$, $i,j=1,2$, $\kappa=\sigma_{11}+\sigma_{22}$, $\hat{A}=\sigma_{11}-\sigma_{22}$,
$\xi=(\frac{1}{4}\hat{A}^2+\mu\sigma_{21}^2)$,  $\Delta\omega=\omega_1-\omega_2$ and $\bar{\omega}=(\omega_1+\omega_2)/2$. The stability of the fixed point $(r_1,r_2)=(0,0)$ is determined by the real part, $Re(\lambda_{\pm})=-\gamma+\frac{\kappa}{4}\pm\frac{1}{2}(\xi)^{\frac{1}{2}}$ for $\Delta\omega=0$, of equation (\ref{tany06}) (that is the fixed points are stable when $Re(\lambda_{\pm})<0$). In this connection we wish to note that when $\sigma_{11}=\sigma_{22}$ and $\mu=1$ the stability equation reduces to the one studied by Montbrio et al. \cite{Montbrio:04} for a system of two interacting populations. In this case the stability diagram for equation (\ref{tany06}) will be similar to Fig. 1 of \cite{Montbrio:04}. When $\sigma_{11}$ increases, the condition $\gamma<\frac{\kappa}{4}\pm\frac{1}{2}(\xi)^{\frac{1}{2}}$ is satisfied and hence the fixed points $(r_1,r_2)=(0,0)$ become unstable. This means that the increase in $\sigma_{11}$ to a sufficiently high value induces synchronization in the target also apart from inducing synchronization in the source.


\subsection{Controlling mass synchronization}

\begin{figure}
\begin{center}
\includegraphics[width=7.50cm]{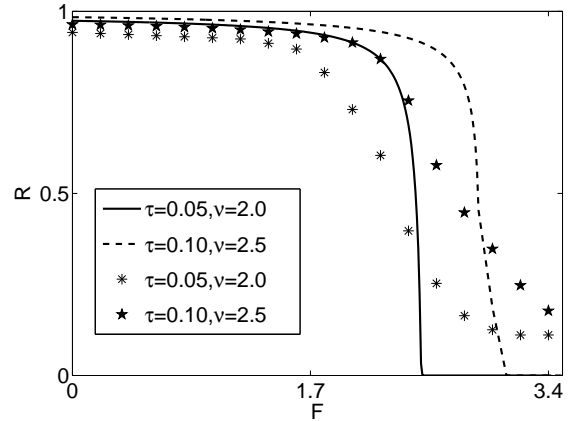}
\caption{The occurrence of desynchronization denoted by a decrease in $R$ due to delayed feedback. The solid and the dotted curves are the analytically obtained solutions given by equation (\ref{cho15}). $\ast$ and $\star$ curves are the corresponding numerical simulations of equation (\ref{cho01}) for $N'=2$, $\gamma=0.02$, $\omega=1.5$ $\sigma_{i1}=\nu$ and $\alpha_{ij}=0$, $i,j=1,2$.}
\label{new_cfi_ana}
\end{center}
\end{figure}

In Sec. 3 we have shown from numerical analysis that pathological synchronization can be controlled by providing a delay feedback. Therefore including the delay feedback term in Eq. (\ref{cho02a}) provides the following amplitude equation,
\begin{eqnarray}
\label{cho11a}
\dot{z}_{\eta}&+&(\gamma_{\eta}-i\omega_{\eta})z_{\eta}+\sum_{\eta'=1}^{N'}\frac{F}{2}
\bigg(z_{\eta'\tau}-z_{\eta'\tau}^{\ast}z_{\eta}^2\bigg)
\nonumber\\&=&\sum_{\eta'=1}^{N'}\frac{\sigma_{\eta\eta'}}{2}
\bigg(e^{-i\alpha_{\eta\eta'}}z_{\eta'}-e^{i\alpha_{\eta\eta'}}z_{\eta'}^{\ast}z_{\eta}^2\bigg),
\end{eqnarray}
where $z_{\eta'\tau}=z_{\eta'}(t-\tau)$. In the pathological synchronization state all the populations are in synchronization with the source and hence the dynamics of all the populations are similar to that of the source and are completely synchronized with it. Thus all the populations have the same synchronization strength and phase, that is $r_\eta\simeq R$ and $\psi_\eta\simeq\phi$, $\eta=1,2,...,N'$. With this reasoning equation (\ref{cho11a}) becomes
\begin{eqnarray}
\label{cho14a}
\dot{R}&=&-\gamma R+(\frac{1-R^2}{2})(\bar{\nu}R
-FR_\tau\cos(\phi-\phi_\tau)),\nonumber\\
\dot{\phi}&=&\omega-(\frac{1+R^2}{2R})(\hat{\nu}R
-FR_\tau\sin(\phi-\phi_\tau)),
\end{eqnarray}
for $\gamma=\gamma_\eta$, $\omega=\omega_\eta$, $\bar{\nu}=\sum_{\eta'=1}^{N'}\sigma_{\eta\eta'}\cos(\alpha_{\eta\eta'})$ and $\hat{\nu}=\sum_{\eta'=1}^{N'}\sigma_{\eta\eta'}\sin(\alpha_{\eta\eta'})$, $\eta=1,2,...,N'$. Here $R$ and $\phi$ are the global order parameters defined by $Z=Re^{i\phi}=\frac{1}{N'}\sum_{\eta'=1}^{N'}r_{\eta'}e^{i\psi_{\eta'}}$. Now with a chage of variable $\phi=\Omega t+\hat{\phi}$ equation (\ref{cho14a}) can be rewritten as
\begin{eqnarray}
\label{cho14ab}
\dot{R}&=&-\gamma R+(\frac{1-R^2}{2})(\bar{\nu}R
-FR_\tau\cos(\hat{\phi}-\hat{\phi}_\tau+\Omega \tau)),\nonumber\\
\dot{\hat{\phi}}&=&\hat{\omega}-(\frac{1+R^2}{2R})(\hat{\nu}R
-FR_\tau\sin(\hat{\phi}-\hat{\phi}_\tau+\Omega \tau)),
\end{eqnarray}
where $\hat{\omega}=\omega-\Omega$. The fixed point $R=0$ of the above equation, which characterizes the fully desynchronized state, is not of physical interest here. So we are now interested in the pathological synchronization state of the system which is represented by the fixed point solution $R\neq 0$ and $\hat{\phi}=$ constant. Consequently from equation (\ref{cho14ab}) we have
\begin{eqnarray}
\label{cho15}
R&=&\sqrt{1-\frac{2\gamma}{\bar{\nu}-F\cos(\Omega\tau)}},\nonumber\\
\Omega&=&\omega+(\gamma-\bar{\nu}
+F\cos(\Omega\tau))\frac{\hat{\nu}-F\sin(\Omega\tau)}{\bar{\nu}-F\cos(\Omega\tau)}.
\end{eqnarray}
For suitable values of $F$ and $\tau$ the strength of the pathological synchronization $R$ can be reduced for a given set of system parameters. This is evident from Fig. \ref{new_cfi_ana} where we have plotted the order parameter $R$ as a function of the strength of the feedback $F$ for two different values of the time delay parameter $\tau$. From Fig. 8, we observe that in the absence of feedback, that is $F=0$, there is a strong synchronization in the system and hence $R\simeq 0.9$. Upon introducing the feedback, the synchronization in the system is destroyed for a suitable values of $F$ for a given $\tau$. The solid and the dotted curves are the analytically obtained solutions given by equation (\ref{cho15}) and the curves denoted by $\ast$ and $\star$ are the corresponding numerical simulations of equation (\ref{cho01}) for $N'=2$. Thus our analytical findings are in reasonable agreement with the numerical results.

Thus we find that the above analytical consideration correlates our earlier numerical study and provides a strong theoretical basis to identify the hypothesis that when strong synchronization occurs in the source it induces synchronization in the target also by increasing the coupling strength between the source and the target. We further conclude that strong synchronization can be controlled by a suitable delayed feedback control.

\section{Conclusion}
\label{conc}
Occurrence of strong pathological synchronization in neuronal populations severely impairs brain function and causes pathological tremors. In general, during pathological synchronization, a particular group of neurons in certain region of the brain gets synchronized and acts as a pacemaker which drives other cortical neuronal populations onto mass synchronization. The synchronization frequency in such cases is close to or equal to the tremor frequency  \cite{Alberts:69,Volkmann:04}.

In this paper we have considered a model of coupled populations of phase oscillators to study the occurrence of strong synchronization. We consider one of the populations to be the source and the other populations that are driven to be the targets. We find the occurrence of pathological synchronization in the model, and from the dynamical properties we are able to hypothesize the possible cause for the occurrence of such strong synchronization: When strong synchronization occurs in the source this in turn increases the coupling strength between the source and the target. In this state the source strongly drives the target and as a result the target gets synchronized with the source. Thus not only the coupling between the source and the target causes the pathological synchronization but also the strength of the synchronization in the source plays a vital role.

We have proposed a possible method to control the occurrence of this pathological synchronization. We find that a delayed feedback given to the system of neuronal populations destroys the strong synchronization and establishes desynchronization. The method is demand-controlled since the strength and the time delay of the feedback is determined from the strength and the frequency of the synchronization. We have analytically analyzed the model in the thermodynamic limit and have deduced the amplitude equation using the ansatz of Ott and Antonsen \cite{Ott:08}. From this equation we have shown the existence of synchronized solutions in the target when strong synchronization occurs in the source. Upon applying delayed feedback we have found that the strength of the synchronization can be reduced. We believe that our hypothesis for the cause of the occurrence of pathological synchronization will be useful for a better understanding and further research in this direction. The proposed method will help in the effective stimulation of neuronal populations in deep brain stimulation therapy.

The work is supported by the Department of Science and Technology (DST)--Ramanna program (ML), DST--IRHPA research project (ML and VKC) and an INSA Senior Scientist fellowship (ML). JHS is supported by a DST--FAST TRACK Young Scientist research project.

\end{document}